\documentstyle[prl,aps,preprint]{revtex}

\begin{document}
\draft
\title{Vortices, flux tubes and other structures in the Ginzburg-Landau 
model:\\ a possible fine structure of the  mixed state?}
\author{Yuri N. Obukhov
\footnote{On leave from: Department of 
Theoretical Physics, Moscow State University, 117234 Moscow, Russia}
and Franz E. Schunck}
\address{Institute for Theoretical Physics, University of Cologne,
 D--50923 K\"oln, Germany}
\date{\today}
\maketitle
\begin{abstract}
\noindent
We present new regular static isolated cylindrically symmetric solutions 
for the Ginzburg-Landau model which have finite Gibbs free energy. These 
configurations (which we call the {\it flux tube} and {\it type B} solutions)
are energetically favorable in the interval of the
external magnetic fields between the thermodynamic critical value $H_{c}$ 
and the upper critical field $H_{c_2}$ which indicates that they 
are important new elements of the mixed state describing a transition from 
vortices to the normal state. 
\end{abstract}


The existence of vortices was predicted by Abrikosov \cite{abr} within the
framework of the phenomenological Ginzburg-Landau (GL) theory of 
superconductivity. Nielsen and Olesen \cite{no}, noticing the mathematical 
equivalence of the GL model to the Abelian theory of coupled gauge and Higgs 
fields, demonstrated the relevance of the vortex (string-like) structures 
in the high energy particle physics. An isolated vortex is a static regular 
cylindrically symmetric solution of the classical second-order GL equations
with the finite energy (line density), see earlier numerical results in
\cite{num}. Our purpose is to present
new exact regular solutions with the {\it finite Gibbs free energy}. We 
consider isolated structures and thus confine ourselves to the case of 
cylindrical symmetry. 

The Lagrangian of the model \cite{no} is
\begin{equation}
L=-{1\over 4}F_{\mu\nu}F^{\mu\nu} - {1\over 2}\left|D_{\mu}\Phi\right|^2 - 
{\lambda\over 4}\left(|\Phi |^2 - 
{\mu^2\over\lambda}\right)^2,\label{lagrNO}
\end{equation}
where $\Phi$ is a complex scalar field, $F_{\mu\nu}=\partial_{\mu}A_{\nu}
-\partial_{\nu}A_{\mu}$ is the electromagnetic field strength, $D_{\mu}=
\partial_{\mu} + iA_{\mu}$, while 
$\lambda$ and $\mu$ are the coupling constants. Since $\mu$ has the 
dimension of an inverse length, it is convenient to introduce for the 
cylindrical system $(\rho, \theta, z)$ a new (dimensionless) radial
coordinate $r$ via $\rho=({\sqrt{\lambda}/\mu})\,r$. Looking for 
{\it static} configurations, we use the cylindrically symmetric 
ansatz $A_{2}=f(r),\;\Phi=({\mu/\sqrt{\lambda}})\,\varphi(r)$,
where $f, \varphi$ are two {\it real} functions. 
Notice that we do not include the phase factor $e^{in\theta}$ for the 
scalar field (cf. \cite{no}). The field $\varphi$ is defined up to a gauge 
transformation and we find it more convenient to work in such a real gauge. 
The magnetic field has only one component in $z$-direction, given by 
$H =({\mu^2/\lambda})\,h$, where we denote the {\it 
dimensionless} magnetic field by $h=f'/r$.

With the above assumptions, the equations of motion reduce to
\begin{eqnarray}
r^2 f'' - r f' &=&r^2\varphi^2 f,\label{1a}\\
r^2 \varphi'' + r\varphi' &=&\varphi\left(f^2 + 
\lambda r^2(\varphi^2 - 1)\right).\label{3a}
\end{eqnarray}

{}From (\ref{lagrNO}), the energy per unit length for static cylindrical
configurations reads
\begin{equation}
{\cal E}={\pi\mu^2\over\lambda}\int\limits_{0}^{\infty}dr r
\left[h^2\! + (\varphi')^2 +
{f^2\varphi^2\over r^2} + {\lambda\over 2}
\left(\varphi^2 -1\right)^2\right].\label{energyNO}
\end{equation}
Fields which satisfy (\ref{1a})-(\ref{3a}) also produce an extremum of 
${\cal E}$. The latter has an absolute minimum ${\cal E}=0$ for the trivial 
solution ($\varphi=1, f=0$) which describes the Meissner state with 
superconducting order in all points of a sample and the magnetic field 
absent inside. The vortex \cite{abr,no,num,super} presents an example of 
a nontrivial finite ${\cal E}$ solution with the magnetic field penetrating
along a thin non-superconducting core around $r=0$. 

Although ${\cal E}>0$ for the vortex, one can prove that it becomes 
energetically more preferable, above the first critical field $H_{c_1}$, than
the trivial Meissner state by inspecting the Gibbs free energy (per unit 
length) which is defined by ${\cal G} = {\cal E} - \int d^2 x\, 
({\hbox{\boldmath $H H$}}_{\rm ext})$. We will assume that the external 
constant magnetic field is also directed along the $z$-axis, and use the 
dimensionless variable defined, as above for $H$,
by $H_{\rm ext}=(\mu^2/\lambda)h_{\rm ext}$. For the understanding of a
transition to the normal state, of particular interest is the functional
\begin{eqnarray}
\Delta{\cal G}&=&{\pi\mu^2\over\lambda}\int\limits_{0}^{\infty}dr r
\bigg[\left(h - h_{\rm ext}\right)^2\nonumber\\
&+& \left(\varphi'\right)^2 +{f^2\varphi^2\over r^2} + 
{\lambda\over 2}\left(\varphi^4 -2\varphi^2\right)\bigg],\label{gibbsdiff}
\end{eqnarray}
which describes the difference ${\cal G}-{\cal G}_{nh}$ between the Gibbs 
free energy of a particular configuration and  that of the normal state 
($\varphi=0, h=h_{\rm ext}$). 
It is important to notice that (\ref{gibbsdiff}) has the {\it same equations 
of extremals} as the energy functional (\ref{energyNO}), namely 
(\ref{1a})-(\ref{3a}). However, unlike the strictly positive ${\cal E}$, 
the functional (\ref{gibbsdiff}) can have any sign. 

Let us study the regular solutions of the GL equations (\ref{1a})-(\ref{3a}).
At the symmetry axis, we find two types of regularity conditions for 
$f$ and $\varphi$:

(A) Potential $f$ is non-zero while scalar field $\varphi$ vanishes 
at the origin,
\begin{eqnarray}
f&=&N + ar^2 + {1\over 4N(N+1)}b^2\;r^{2N+2} + 
O(r^{2N+4}) ,\label{ser1}\\
\varphi&=&b\,r^N\left(1+ {N\over 2(N+1)}\left(a-
{{\lambda\over 2N}}\right)r^2 + O(r^4)\right),
\label{ser3}
\end{eqnarray}
where $N=\pm 1,\pm 2,...$ and parameters $a,b$ are arbitrary.

(B) Potential $f$ vanishes while scalar field $\varphi$ is nontrivial at the 
origin,
\begin{eqnarray}
f&=& ar^2\left(1 + b^2\;r^2/8 + O(r^4)\right) ,\label{0ser1}\\
\varphi&=&b\,\left(1+ 
{\lambda}(b^2 - 1)r^2/4 + O(r^4)\right),\label{0ser3}
\end{eqnarray}
with some parameters $a,b$.

The vortex solution satisfies (A) and is distinguished by the well known
asymptotic conditions at infinity ($\varphi\rightarrow 1, f\rightarrow 0$).
Analysing the (B) case, one can prove that {\it no finite 
energy} solutions for type B regularity conditions exist. However, it is
possible to find {\it finite Gibbs free energy} regular type B solutions 
which satisfy at infinity
\begin{eqnarray}
h(r)={1\over r}f'(r)\,\vline\,{\hbox{\raisebox{-1.5ex}{\scriptsize 
$r\rightarrow\infty$}}}
&&\longrightarrow h_{\rm ext},\label{infpotB}\\
\varphi(r)\vert_{r\rightarrow\infty}&&\longrightarrow 0.\label{infscalB}
\end{eqnarray}

The simplest solutions with $\varphi$ having no nodes are shown in 
Fig.~\ref{fig1} (in all figures we use $\kappa:=\sqrt{\lambda}$). There 
exist also more nontrivial solutions with nodes. A principal difference of 
these solutions (which we will call ``type B'' solutions for brevity) from 
the vortex solutions with a quantized flux lies in the fact that the 
magnetic field is now asymptotically constant, and hence the total magnetic 
flux is infinite. However a reasonable replacement of the flux is provided 
by 
\begin{equation}
M:= \int \rho d\rho d\theta(H-H_{\rm ext}) = 
2\pi\int\limits_{0}^{\infty}dr r (h - h_{\rm ext}).\label{diflux}
\end{equation}
Defined formally as a difference of fluxes, this variable is usually
interpreted as a {\it magnetization} per unit volume \cite{super}.
Unlike the quantized flux for the vortices, $M$ can have an arbitrary value
for the type B solutions. 

Numerical integration reveals the following interesting properties of the 
nodeless type B solutions: For $\lambda >1/2$ they all have 
$\Delta{\cal G}<0$, for $\lambda =1/2$ always 
$\Delta{\cal G}=0$, and $\Delta{\cal G}>0$ for $\lambda <1/2$.
Hence, for an ideal type II superconductor the type B configuration is 
energetically more preferable than the normal state. However, 
the type B solutions do not exist for every value of the external field
$h_{\rm ext}$. For a given $\lambda >1/2$, one finds $h_{\rm ext}\leq\lambda$
for all solutions, and thus the upper limit is exactly equal to
the {\it second critical field} $h_{c_2}=\lambda$.

However, this is not the end of the story. Besides the type B configurations,
there are other nontrivial solutions with finite values of the Gibbs 
functional (\ref{gibbsdiff}). Namely, let us take the type A regularity 
conditions (\ref{ser1})-(\ref{ser3}) at the origin, while at infinity we 
consider the asymptotics (\ref{infpotB})-(\ref{infscalB}), which 
suggests a possible physical interpretation of such solutions as emerging 
from a ``gluing'' of a vortex with a type B configuration. 
These solutions are shown in Fig.~\ref{fig2}. Their form explains why we 
call them {\it flux tubes}~\cite{foot2}: There is a non-superconducting core 
filled by the magnetic field (looking like a vortex), surrounded by a 
superconducting tube (almost completely free of the magnetic field). Outside 
such a tube the sample quickly reduces to a normal state ($\varphi=0$) 
with a constant external field in it. 

Each family of the flux tube (FT) solutions has two branches. For the 
nodeless $\lambda=1$ case, we find that one of these branches is 
characterized by the positive Gibbs free energy, and another has negative 
Gibbs free energy. However, for $\lambda >9/2$ both branches describe 
configurations with 
$\Delta{\cal G}<0$. We find it convenient to depict these 
branches in the form of magnetization curves, see Fig.~\ref{fig3}. For 
$\lambda\leq 1/2$, both branches have $\Delta{\cal G}>0$.

Similarly to the type B configurations, we find the flux tube solutions 
with nodes. For $\lambda=1$, all the FT solutions with one node have 
$\Delta{\cal G}>0$. However, with increasing $\lambda$ this changes. The 
crucial point is the position of the thermodynamic critical value $h_{c}=
\sqrt{\lambda/2}$ relative to the ``limiting values'' of $h_{\rm ext}$ at 
which a FT family has vanishing magnetization and Gibbs free 
energy. Numerically, we discovered that these limiting points are located, 
on the $h_{\rm ext}$ axis, at $\lambda, {1\over 3}\lambda, {1\over 5}\lambda, 
{1\over 7}\lambda, \dots$. Below we explain these values with the help of 
the linearization analysis. 

Since both, the type B and the FT solutions, have a well defined
finite Gibbs free energy, it is natural to compare them. We find that 
always $\Delta{\cal G}_{\rm FT}<\Delta{\cal G}_{\rm B}$. 
For illustration, we display $\Delta{\cal G}$ as a function of $h_{\rm ext}$
for $\lambda=(2.25)^2$ in Fig.~\ref{fig4} which shows that 
{\it the flux tube configurations are energetically more  preferable}. 
This holds true for any $\lambda$.

The best way to understand the structure of the type B and FT solutions 
in the limit of vanishing magnetization ($M\rightarrow 0$) and Gibbs free
energy ($\Delta{\cal G}\rightarrow 0$) is to study the linearized 
GL equations. The linearization of the system (\ref{1a})-(\ref{3a}) leads to
the Schr\"odinger-type equation for $\varphi$ with the potential of 
a circular oscillator. Regular solutions exist only when 
\begin{equation}
h = {s\lambda\over {1 + 2n + sN + |N|}},\label{eigenvalue}
\end{equation}
where $n=0,1,2,...\,$ and $s=\pm 1$. Corresponding eigenfunctions 
$\varphi_{n,N}$ are given in terms of the Laguerre polynomials, with $n$ 
equal to the number of their zeros (nodes). Let us denote $h_{k}:=
\lambda/(2k+1), k=0,1,2,\dots\,$. For $n=0$ and $sN=-|N|$, one finds that
the maximal eigenvalue of (\ref{eigenvalue}) is $h=s\lambda=sh_{0}$,
which is precisely 
the second critical field $h_{c_{2}}=\lambda$. At the first sight, this fact 
may seem to be well known in the literature on the type II superconductors,
cf. \cite{abr,super}. However, the peculiar point is the conclusion that the 
vortices apparently have nothing to do with the determination of 
$h_{c_{2}}$. As we see, the latter is determined by the existence condition 
for the nodeless type B and FT solutions which are energetically more 
preferable than the normal state. It is clear that the eigenfunctions 
$\varphi_{0,N}=r^{|N|}\exp(-{\lambda\over 4}r^2)$ describe the linearized 
regular type B (for $N=0$) and FT (for $N\neq 0$) solutions. It seems 
worthwhile to notice also that the starting point for the well known 
(see, e.g., \cite{abr,super}) construction of a vortex lattice near $h_{c_2}$ 
is not the vortex, as one might think, but a linearized type B solution. 

The physical meaning of the lower eigenvalues (\ref{eigenvalue}) is now
also clear: these define the limiting values for the external magnetic field 
at which the exact type B and FT solutions become ``linearizable'' and thus 
disappear with vanishing magnetization and Gibbs free energy. This 
confirms our numerical observations: For instance, both the nodeless FT and 
the type B configurations have the limiting magnetic field values 
$h_{1}=\lambda/3$ and $h_{0}=h_{c_2}$, while the FT and the type B with 
one node ``live'' between $h_{2}=\lambda/5$ and $h_{1}=\lambda/3$. 
In general, the limiting points for the solutions with $k$ nodes are
$h_{k+1}$ and $h_{k}$. As we already mentioned, for a given $\lambda$ the 
position of the thermodynamic critical value $h_{c}=\sqrt{\lambda/2}$ 
relative to the interval $[h_{k+1}, h_{k}]$ is important.
If $h_{c}$ belongs to this interval, then 
$h_{k+1}<h_{\rm ext}<h_{k}$ for all solutions in
this family. However when $h_{c}$ does not belong to this interval,
then $[h_{k+1}, h_{k}]$ is extended up to $h_{c}$. In particular, we 
find that for every FT family, the magnetization diverges when $h_{\rm ext}
\rightarrow h_{c}$. Moreover, one can show that the Gibbs free energy for a 
vortex and for a flux tube solution are equal for $h_{\rm ext}=h_{c}$,
\cite{our}.

One may further weaken the conditions at infinity, requiring $f\rightarrow 0$
for the potential, but for the scalar field simply demanding regularity 
and finiteness. This gives another class of regular configurations which we
call the oscillating solutions (OS). For large $r$ we can neglect the 
nonlinear terms in (\ref{3a}), and get for the scalar field the linearized 
equation $\varphi'' + \varphi'/r + \lambda\varphi =0$. Its solution is the 
Bessel function $\varphi=J_{0}(\sqrt{\lambda}r)$ which is confirmed by 
the direct numerical integration. Such OS seem to be
relevant to observations made in \cite{fin,our}. In our opinion these new 
solutions are unstable configurations preceding to the completely formed
Abrikosov vortices. Notice, that an oscillating character of such
solutions may resemble an ``intermediate'' superconductor state with
coexisting normal and superconducting regions. But a big 
difference is that the magnetic field penetrates only at the center, and the
magnetic flux is quantized precisely as for the vortices.
Oscillating solutions exist both for the type A and B conditions,
in the latter case the magnetic field is completely zero. 
A curious feature is that for a finite sample in absence of an external 
magnetic field an oscillating state is energetically more preferable
than a purely normal state. 

In each class of solutions the decisive role is played by the values of
the parameters $(a,b)$ appearing in the regularity conditions 
(\ref{ser1})-(\ref{ser3}) and (\ref{0ser1})-(\ref{0ser3}). Numerically,
all the reported new solutions were obtained after a ``fine tuning'' of 
these parameters. Let us draw a kind of a ``phase diagram'' 
on the $(a,b)$ plane showing explicitly the domains of existence for 
different solutions. Since the vortices, flux tubes, and oscillating 
solutions all satisfy the type A conditions (\ref{ser1})-(\ref{ser3}), 
we can display them on the same $(a,b)$-plane, see Fig.~\ref{fig5} for
the case of $\lambda=1$. The big dot shows the ``position'' of a vortex,
while each curve represents a complete family of 
solutions (a point on a curve gives $(a,b)$ for
a particular solution). The curves hit the $a$-axis ($b=0$) at one of 
the possible limiting values: $a=\pm{1\over 2}h_{k}, k=0,1,\dots$. Curves 
for FT with increasing number of nodes are ``concentrating'' around 
the oscillating curve which seems to indicate that the oscillating 
solutions are unstable and a small perturbation may cause their decay into 
a nearby flux tube with a finite number of nodes. Moving along any FT
curve away from the $a$-axis, one hits the vortex dot, where
magnetization diverges. 

In this paper we have presented new numerical solutions of the
cylindrically symmetric GL equations. Noticing that the latter are common
extremals for the energy (\ref{energyNO}) and the Gibbs free energy 
(\ref{gibbsdiff}) functionals, we describe the regular solutions 
with {\it finite Gibbs free energy}. We show that $h_{c_2}$ is not the
magnetic field below which the vortex becomes energetically more preferable 
than the normal state, but it is the field at which the formation of 
the type B and flux tube structures starts, since 
$\Delta{\cal G}_{\rm FT,B}<0$ for $h_{\rm ext}<h_{c_2}$. 
Our results show that the flux tube solutions without node remain  
energetically most preferable from $h_{c_2}$ down to $h_{c}$, after 
which the vortices become energetically dominant.
It is worthwhile to stress, that our results do not contradict the previous 
knowledge about the mixed state in the type II superconductors. Notice that, 
after all, one can interpret a flux tube solution as a vortex ``surrounded'' 
by a type B configuration. 
In a certain sense, one can speak of a ``fine structure'' for the mixed
state: Different configurations (vortices, type B and FT solutions with 
$k$ nodes, $k=0,1,\dots$) can exist for any external field $h_{\rm ext}$ 
between $h_{c_1}$ and $h_{c_2}$ (cf. Table \ref{table}). 

We are grateful to Friedrich W. Hehl and Eckehard W. Mielke for extremely
helpful advice and discussions. The work of Y.N.O.  was supported by the 
Deutsche Forschungsgemeinschaft (Bonn) grant He 528/17-1, and for F.E.S. by 
the European Union.

\begin{figure}
\caption[]
{Type B solutions without node for $\kappa =1.0$: The scalar field $\varphi$
and the magnetic field $h$ with initial values
$\varphi (0)=0.9, 0.6, 0.3$ and $f(0)=0.0$.}
\label{fig1}
\end{figure}

\begin{figure}
\caption[]
{Flux tube solutions without node for $\kappa =0.5, 1.0, 1.5$ with the
fixed value of magnetization.}
\label{fig2}
\end{figure}

\begin{figure}
\caption[]
{Magnetization curves for nodeless FT with 
$\kappa =0.5$, $1.0, 1.5$. For each $\kappa$, the corresponding limiting 
values $h_1$, $h_2$, $h_{c}$, and 
$h_{c_2}$ are drawn. The magnetization diverges at $h_{c}$.}
\label{fig3}
\end{figure}

\begin{figure}
\caption[]
{$\Delta {\cal G}$ against the external magnetic field for
the flux tubes and the type B solutions with $\kappa =2.25$. The flux tubes 
have lower $\Delta{\cal G}$ and thus are energetically more preferable.} 
\label{fig4}
\end{figure}

\begin{figure}
\caption[]
{$(a,b)$-diagram for $\kappa =1.0$. The big dot describes
the corresponding vortex solution. The drawn, broken and dotted 
lines represent the nodeless flux tubes, flux tubes with one node, 
and oscillating solutions, respectively.}
\label{fig5}
\end{figure}

\pagebreak
\input psfig

\centerline{FIGURE 1:}
\centerline{\psfig{figure=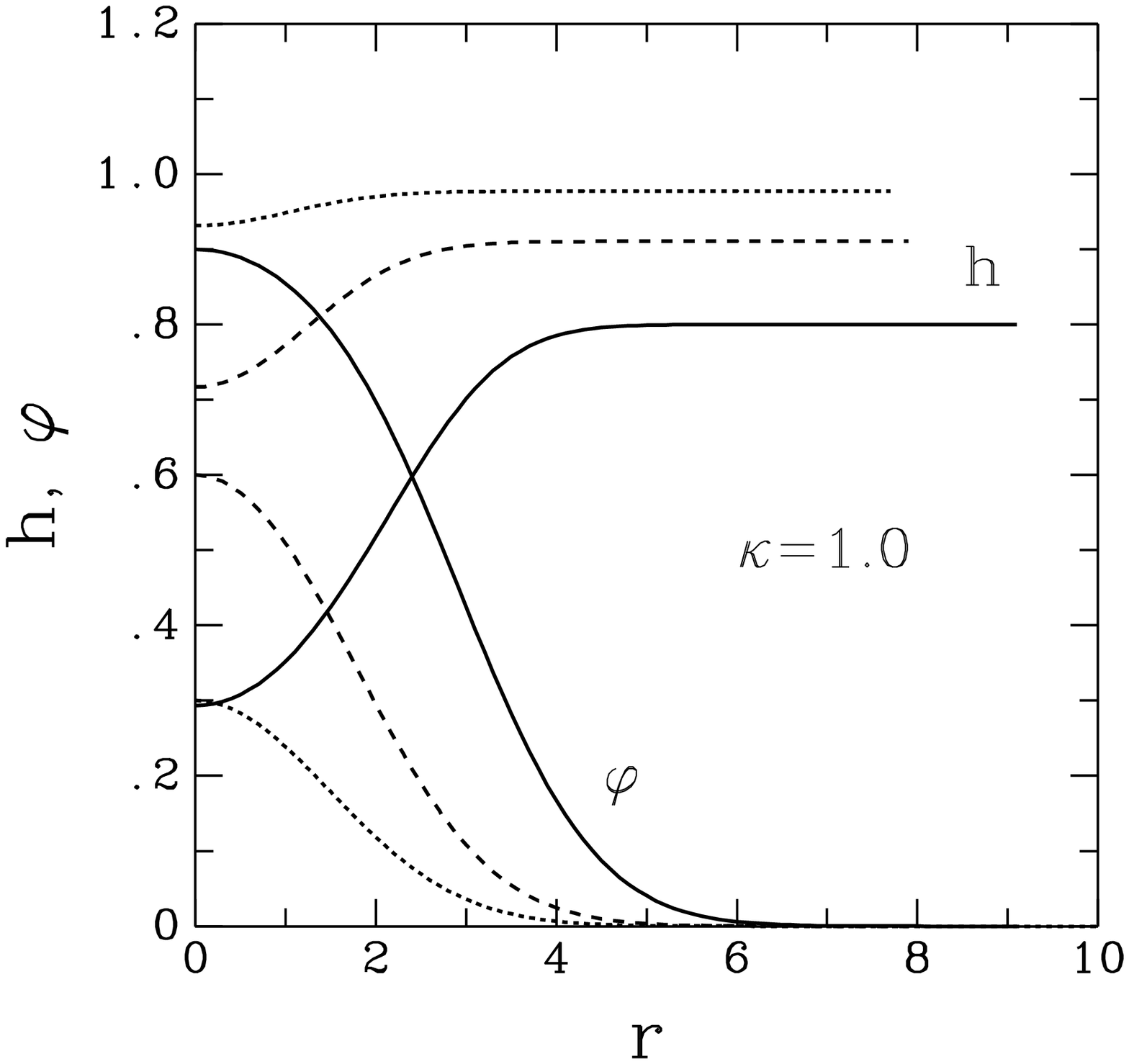,height=6cm}}
\centerline{FIGURE 2:}
\centerline{\psfig{figure=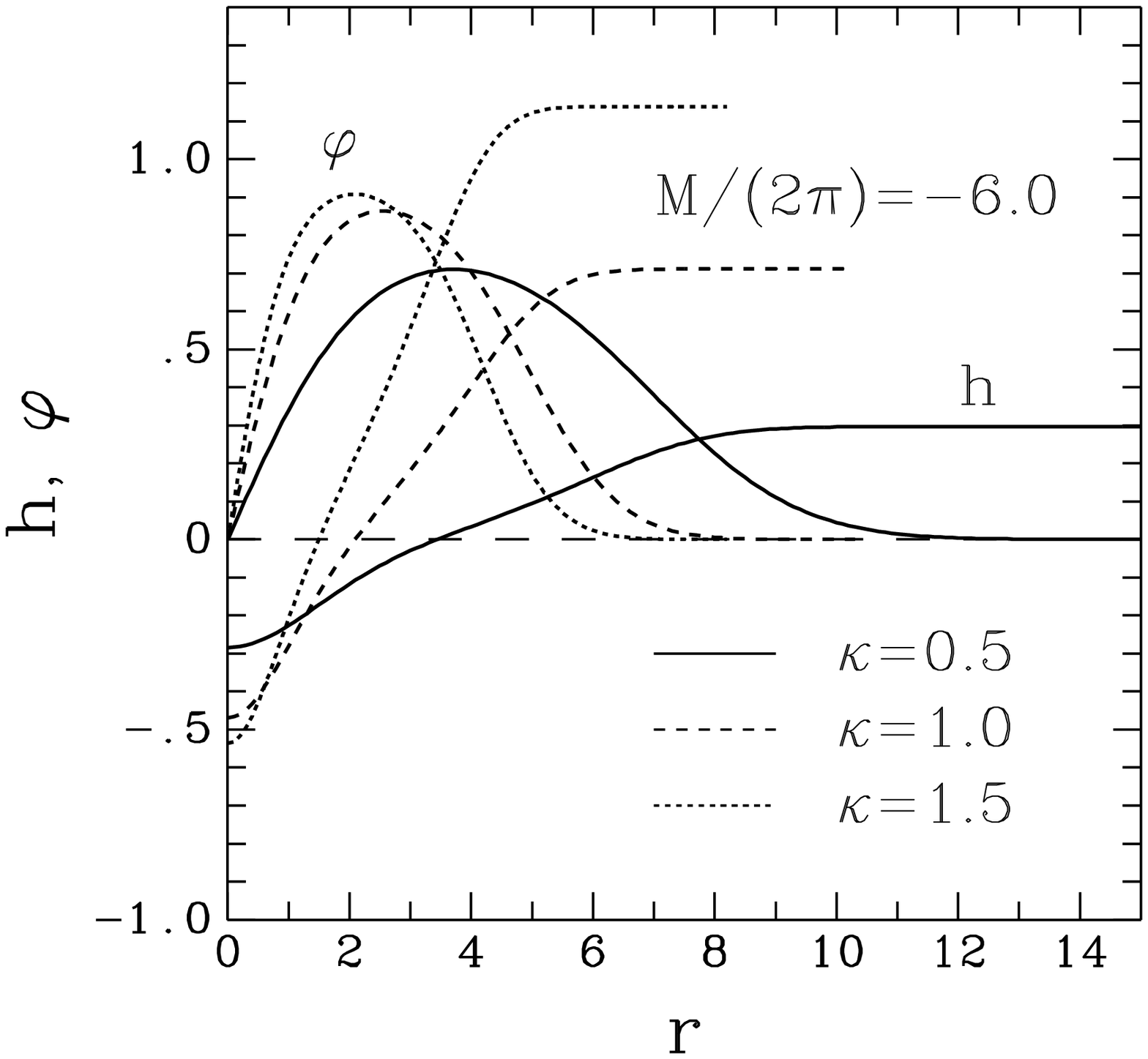,height=6cm}}
\centerline{FIGURE 3:}
\centerline{\psfig{figure=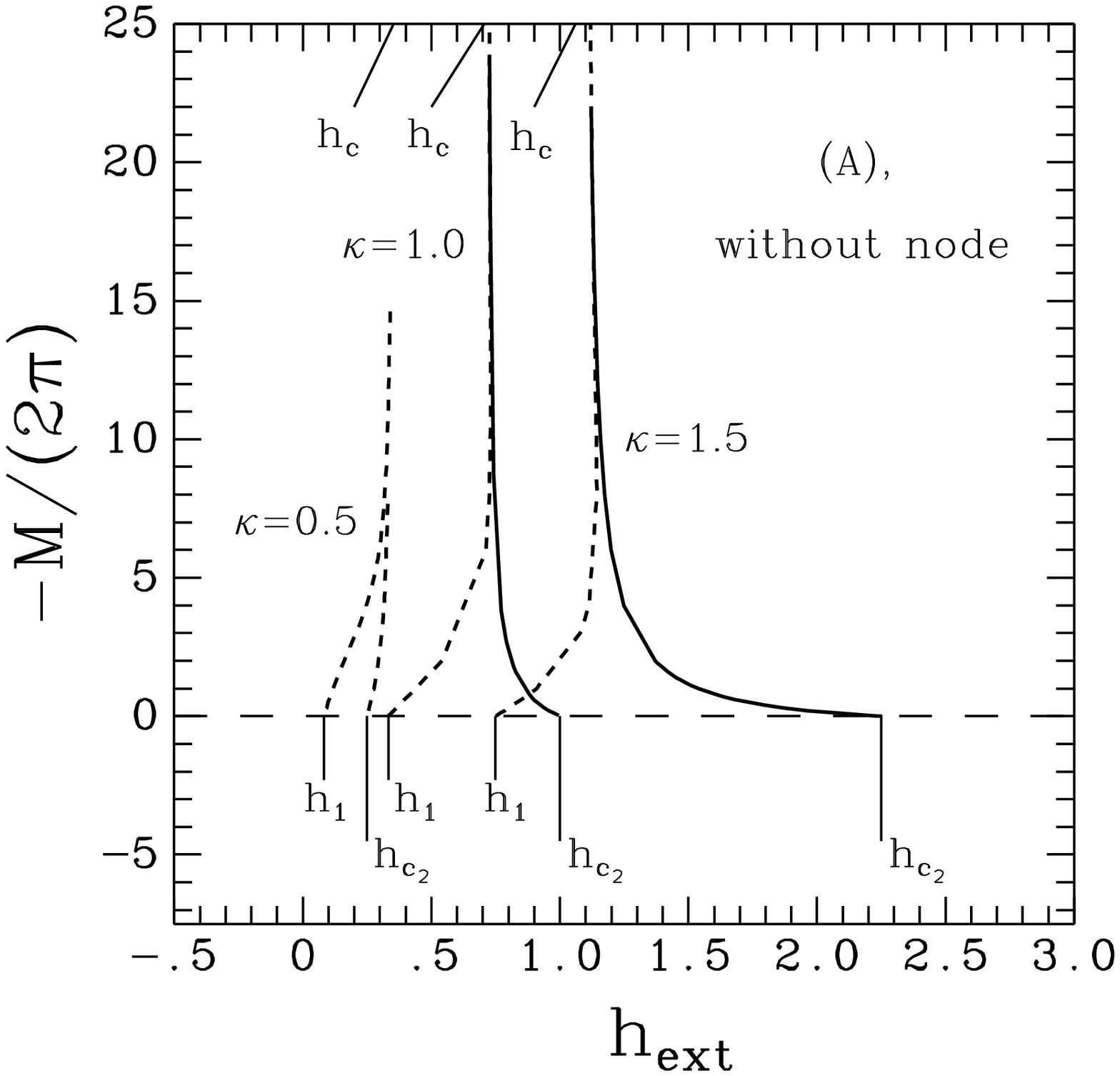,height=6cm}}
\pagebreak
\centerline{FIGURE 4:}
\centerline{\psfig{figure=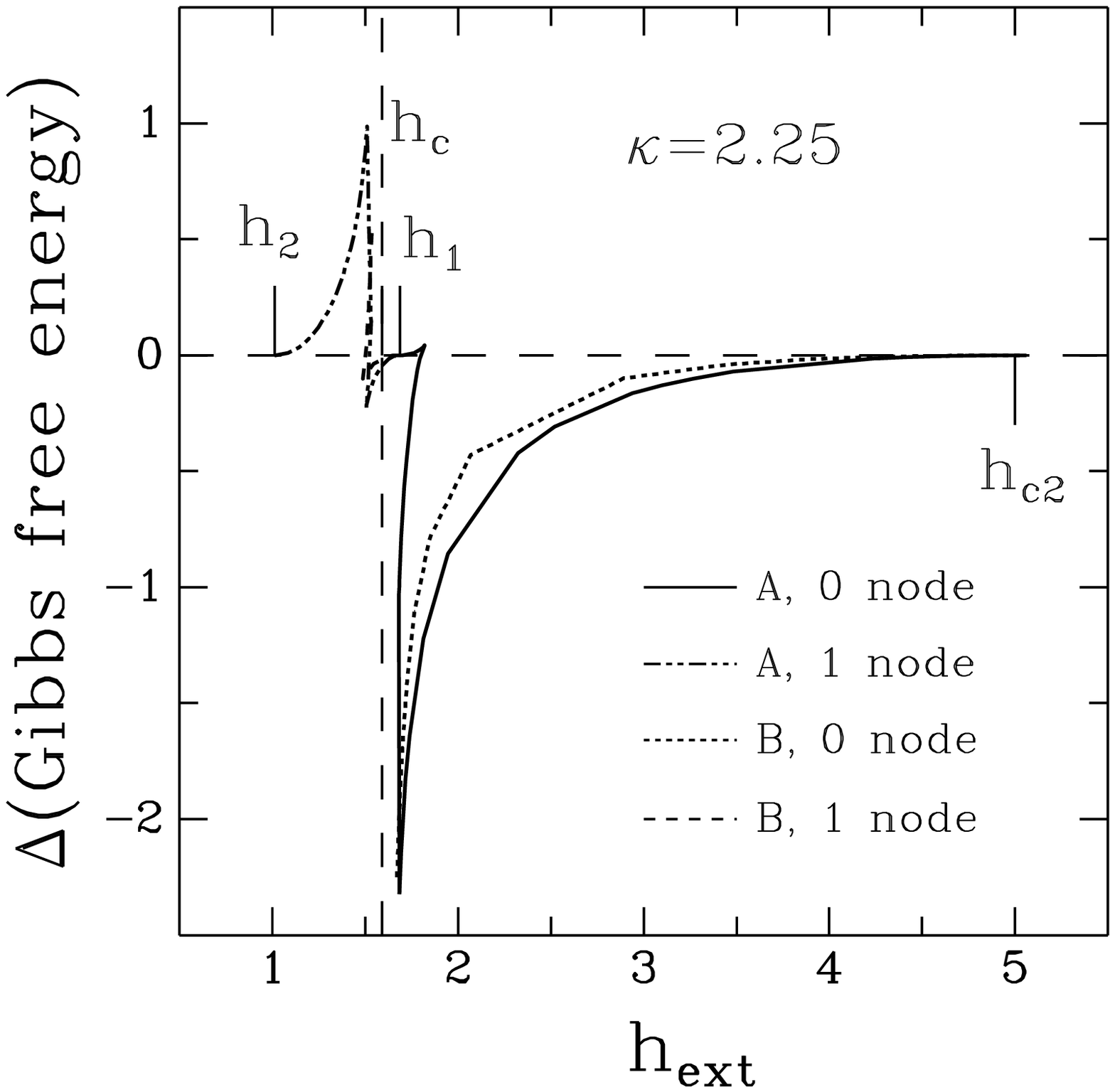,height=6cm}}
\centerline{FIGURE 5:}
\centerline{\psfig{figure=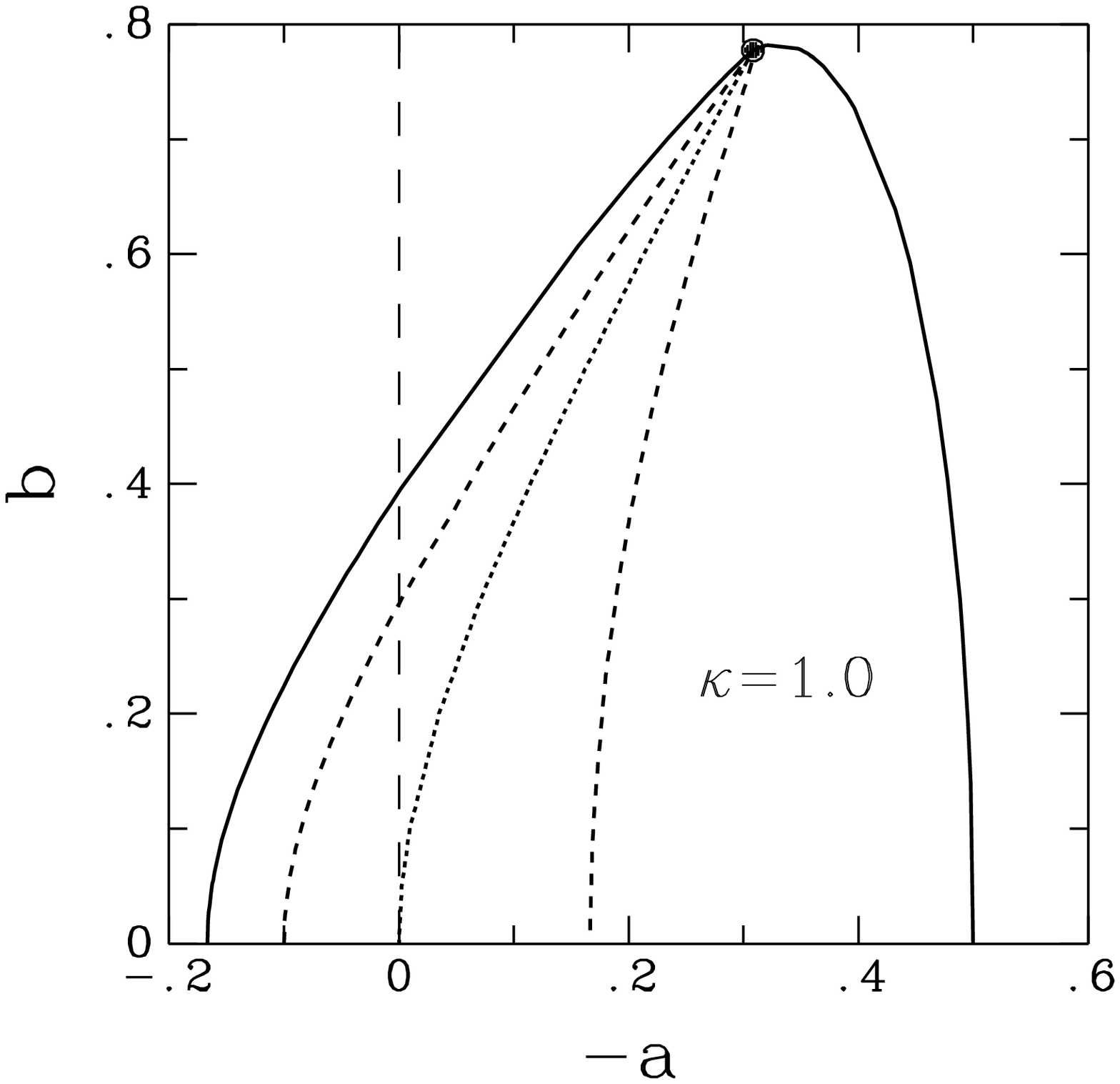,height=6cm}}

\begin{table}
\caption[]{Existence domains of different GL solutions: 
fine structure of the mixed state?}
\begin{tabular}{ll}
region & energetically preferable \\
\tableline
$h_{c_3}<h_{\rm ext}$           & normal \\
$h_{c_2}<h_{\rm ext}<h_{c_3}$   & surface superconductivity \\
$h_{1}  <h_{\rm ext}<h_{c_2}$   & 0-node FT \\
$h_{2}  <h_{\rm ext}<h_{1}$     & 0-node FT, 1-node FT \\
$h_{3}  <h_{\rm ext}<h_{2}$     & 0-node FT, 1-node FT, 2-node FT \\
$h_{c_1}<h_{\rm ext}<h_{c}$     & pure vortices \\
$0      <h_{\rm ext}<h_{c_1}$   & Meissner \\
\end{tabular}
\label{table}
\end{table}

\end{document}